\documentclass{acm_proc_article-sp}

\begin{document}
%

\title{Do Retweets indicate Interest, Trust, Agreement?\\
(Extended Abstract)}
%
%
%
%
%

\numberofauthors{1} 
%
\author{
%
%
\alignauthor
P. T. Metaxas, 
E. Mustafaraj, 
K. Wong, 
L. Zeng, 
M. O'Keefe, 
S. Finn 
\\
       \affaddr{Department of Computer Science}\\
       \affaddr{Wellesley College, Wellesley, MA, USA}\\
}

\maketitle
\begin{abstract}
Arguably one of the most important features of Twitter is the support for ``retweets'' or  messages re-posted verbatim by a user that were originated by someone else. 
(This does not include modified tweets that sometimes are referred to as retweets.)
Despite the fact that retweets are routinely studied and reported, many important questions remain about user motivation for their use and their significance. 
In this paper we answer the question of what users indicate when they retweet. We do so in a comprehensive fashion, by employing a user survey, a study of user profiles, and a meta-analysis of over 100 research publications from three related major conferences.  
Our findings indicate that retweeting indicates not only interest in a message, but also trust in the message and the originator, and agreement with the message contents. However, the findings are significantly weaker for journalists, some of whom beg to differ declaring so in their own user profiles. On the other hand, the inclusion of hashtags strengthens the signal of agreement, especially when the hashtags are related to politics.
While in the past there have been additional claims in the literature about possible reasons for retweeting, many of them are not supported, especially given the technical changes introduced recently by Twitter.

\end{abstract}




\section{Introduction}
\label{Introduction}

Twitter is a real-time information network
that allows its users to write short messages (``tweets'') up to 140 characters in length. 
We chose to focus on Twitter in particular because, unlike other popular social media platforms, it has grown into a real-time news source created by everyday users. Twitter is credited for its role in political events such as monitoring elections, aiding the so-called  ``Arab Spring'' \cite{bruns2013arab}, and for drawing attention to news stories that were initially largely ignored by traditional news media such as the Wendy Davis filibuster\footnote{Texas filibuster on abortion bill rivets online, by Heather Kelly. CNN.com, June 26, 2013. 
}  and the Michael Brown shooting in Ferguson, MO. 


Over the years Twitter has developed its own syntactic components. 
This paper focuses primarily on retweeting:
Users {\bf retweet (RT)} to 
forward a message from another source to their own followers.  Retweeting was not part of the original design of Twitter, but was created through user initiative. 
Because of its later adoption date, tweet forwarding can be done in one of two ways (one that is supported by the Twitter API and another that is not): either by clicking a ``Retweet'' button provided by the Twitter client (and some other clients), or by modifying the original tweet and manually typing ``RT @[user]''  or adding ``via @[user]'' in a new tweet. We call the tweets produced by typing ``modified retweets'' (MRTs) to distinguish them from the first kind. This distinction is important because an unmodified retweet is treated differently by the Twitter platform and is guaranteed to point to its originating source.


The retweet button's utilization initially was slow. In 2010,
\cite{SHPC10} counted 2.99M retweets generated by the retweet button on Twitter, which was 36.34\% of the total of 8.24M retweets collected in a corpus of 74M tweets.  
In a collection of 39M tweets in mid-summer of 2012, about one fifth of retweets are created manually by users.


As a platform, Twitter has several characteristics that make it convenient for research. 
Yet, it is a valid question to ask whether it is interesting to study any specific social media platform. If Twitter does not exist in a few years, will this paper's findings matter?
We argue that the answer is yes: Twitter is recording human {\em communication} that requires relatively little effort to produce and consume. This communication is what we want to study.
While any particular social media platform may cease to exist or lose popularity in the future, the importance of human interaction through social media is unlikely to change. Humans are social animals and their desire to communicate with each other and comment on their social environments is one of their universal and unique characteristics. 
We have chosen to study the interaction of humans through social media in an abstract way, not a way specific to the particular social media platform. We are simply looking at the behavior as revealed through their interactions. 
Further, 
its service makes it easy for people to say something. The effort in contributing to the general social dialog is far less than that of writing a comment  a blog on a web site or a newspaper op-ed, and it has wider impact than talking person-to-person or via email. In addition, the effort to propagate a message sent by someone else is also remarkably small -- giving rise to degrading characterizations of online social participation such as ``slactivism'' \cite{G10}.

So, why do people retweet? The published literature does not give a conclusive answer to this question, though it was studied as early as 2010 \cite{BGL10}. Some of the claims made in the past can be characterized as straightforward (in the sense of being self-evident, ie. that people retweet to broadcast information) and others as outdated (in the sense that technical changes in Twitter do not support them anymore, ie. that people retweet to appropriate information).
However there is no agreement in the literature whether retweeting indicates trust, agreement or even endorsement of the originator or the message. Moreover, there is no clear understanding how the retweeting rates are affected by emotion, political intention, or propaganda, if they are affected at all. Our paper tries to answer conclusively these questions. 

The remainder of this extended abstract is organized as follows: In the section~\ref{survey} we present the results of the user survey that answers the question of what Twitter users think about their practice of retweeting. In subection~\ref{reporters} we study an important user subgroup, reporters, that includes members whose retweeting practice appears to differ significantly from that of the general user. We also present an analysis of user profiles that include a disclaimer about retweeting not indicating endorsement. Section~\ref{meta-analysis}, truncated in this short paper, presents a meta-analysis of research papers from three major conferences related to Twitter data. Finally, section~\ref{conclusions}
has our conclusions.

\section{User Survey}
\label{survey}

In mid-July 2014, we conducted a survey asking Twitter users about various aspects of their retweeting behavior. The survey responses indicated that there is general consensus on several important factors that the majority of Twitter users consider when deciding whether or not  to retweet a given message. However, we also found that there were also significant variations across subcategories of users, including differences across the responses from reporters and non-reporters,  individuals interested in politics, users of different ages, and users who use Twitter with different levels of frequency. 

\subsection{Survey Methodology}

We designed a comprehensive twenty-one item questionnaire that asked participants about their retweeting behavior. Initial survey questions inquired about whether the participant had a Twitter account, and if so, for how long. Other questions relating to general Twitter usage included how often the participant uses Twitter, how often they tweet, why they use Twitter, how many followers they have, and how many accounts they follow. The survey also asked participants about what factors they find important when deciding whether or not to follow an account on a scale from 1 (not at all) to 5 (extremely). 

After answering these questions, we asked participants to indicate how often they retweet, if at all. Only participants who responded that they are familiar with retweeting were led to a page of questions asking them about what factors, both about the message and the account,  they consider important when choosing to retweet a post. In terms of the message, options were whether they find the message interesting, emotionally resonant, entertaining, something they endorse, something that they agree with, something their followers might find interesting, something formative or something they find trustworthy. In terms of the account, options were whether they find the account credible, a celebrity, someone they know personally, someone who shares their opinion, someone they support, or someone they trust. Both  cases were multiple-choice ending with a write-in (``other'') option. These questions were posed as statements on a likert-scale from 1 (strongly disagree) to 5 (strongly agree). They were also asked about what is more important when they decide to retweet: the message, the account or both. 

Additionally, participants who had reported retweeting in the past week were asked to recall certain features about one particular retweet of their choice. These additional questions were designed to demonstrate whether users' actual behavior is consistent with their overall impressions of their own behavior. The final questions captured basic demographic questions about the participants including age, self-described ethnicity, gender and educational level.

\subsection{Survey Results}
We launched the study by utilizing our social links and reached directly audiences, primarily in academic and ethnic communities that we had social and professional links (Twitter, Facebook, Linkedin and email). 
Reflecting on the mode of launching our survey, the responders' sample ($n$=316) was highly educated, with 25.3\% possessing a Bachelor's degree and an additional 43\% a Master's Degree or higher. Of the responders, 
thirty-three (11.4\%) identified themselves as bloggers, journalists, or reporters.

Our sample was representative of both males (43\%) and females (54.4\%), with 2.6\% chosing not to disclose gender. In terms of ethnicity, we left it as an open-ended question and recoded participants responses as  Caucasian/White (46\%), Asian or Pacific Islander (11\%), Black (3.8\%), other (22\%), and no-repyling (17\%). 

\begin{figure}[ht] 
\centering
\includegraphics[width=0.5\textwidth]{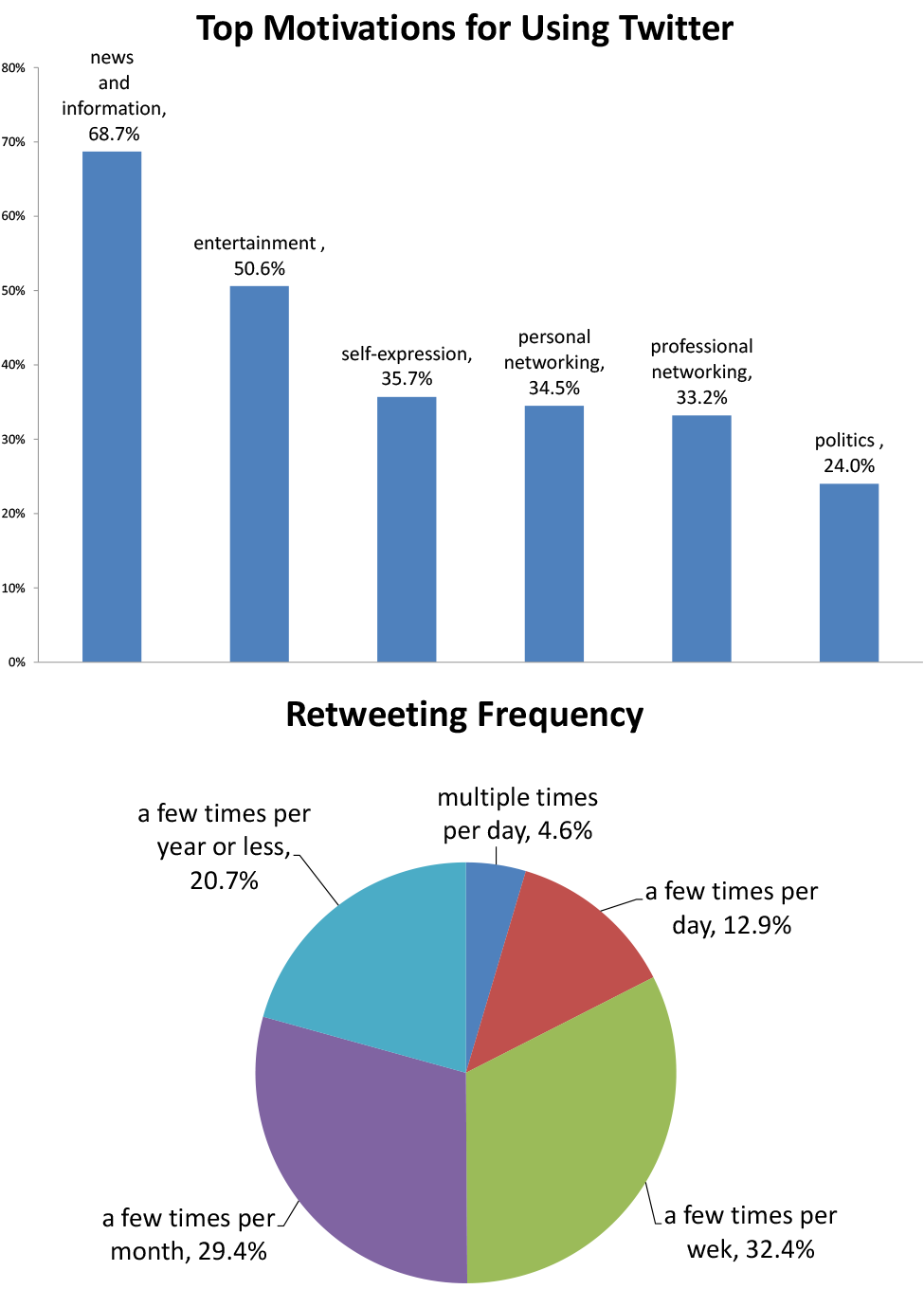}
\caption{\label{fig:charts} Summary of the results of our survey why people use Twitter (top), and how often do they retweet (bottom).  
}
\end{figure}

The age demographics of our data is consistent with Pew's report of Twitter users\footnote{Pew Research Internet Project: Social Networking Fact Sheet. Data from 2014. 
} on several respects. Reflecting the similar trend that younger individuals tend to be on social media platforms more than older ones, including Twitter, 38.6\% of the participants report being under the age of 24, 25.9\% between 25-34, 13.9\%  between 35-44, 12.3\% between 45-54, and 9.2\% over the age of 54. 
Additionally, our data captured users who use Twitter on average as 42.2\% of participants use Twitter daily, 23.8\% use it weekly, and 34\% use it monthly or less. 
55.7\% have less than one hundred followers while 8\% have over one thousand.
40\% of participants provided their Twitter handle.
Figure~\ref{fig:charts} summarizes  some of our results about participants primary motivations for using Twitter and their retweeting behavior.    

When it came to retweeting behavior, 
the majority of participants responded that they believed that {\em the message was more important than the account} when choosing to retwteet a post.  
That is, when chosing whether or not to retweet a tweet, they evaluated and put more significance on the textual content of the tweet than on the user who had written the tweeted. In fact, some users even reported to not pay attention to who tweeted the post in the first place.

\begin{figure*} [!h]
\centering
\includegraphics[width=0.9\textwidth]{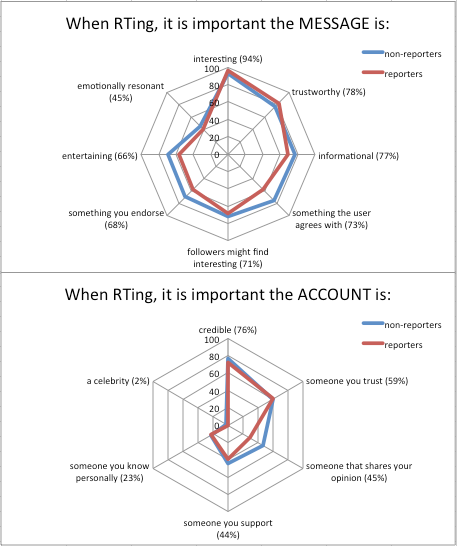}
\caption{\label{fig:message-account-combined-radial} 
(top) 94\% of survey participants indicate that personal interest in a message is the primary reason for retweeting. Reasons that get very strong support (about 3 out of 4 users) are trustworthiness, informativeness and user agreement with a message. More 2 out of 3 users also indicate as reasons to retweet curation (interest to their followers), message endorsement and entertaining nature of the message. Emotional resonance with the message was not viewed as an important reason.
(bottom) Participants indicated that account credibility (76\%) and trustworthiness (59\%) are the most important factors about the account were most important when deciding to retweet a post. Sharing opinion with and support of the originator were not viewed as important reasons, especially for reporters. Little or no support has personal knowledge of the originator and whether the originator was a celebrity.
}
\end{figure*}

Participants provided consistent responses when responding to survey questions that asked about various factors they considered important about a given message when choosing whether or not to retweet it. The majority of participants indicated that factors such as how interesting the message is to themselves as well as their followers are important.  They also cited trustworthiness, informativeness, and whether they agree with the given message as  other factors that were important (see Fig.~\ref{fig:message-account-combined-radial}, top).



Some of the initial insights that we found helped guide our further exploration of the data that was collected. First, we found 
that the majority of participants tended to retweet posts from accounts that they follow (as opposed to accounts that they do not follow) and therefore it is important to understand why they choose to follow certain accounts in the first place.

Consistent with the hypothesis that users care about credibility and personal interest when navigating through social media platforms for information, responses to the survey indicated that specific factors including ``shares my interest" and ``trustworthiness" were most commonly cited as important factors to consider when choosing whether or not to retweet a given account (see Fig.~\ref{fig:message-account-combined-radial}, bottom). Meanwhile, factors such as ``celebrity" and ``entertainment" were not rated as important in Twitter users' decision-making process. Additionally, there were other factors that yielded responses similar to that of ``shares my opinion" where there was no strong opinion. The participants' responses were recoded from a likert-scale from 1 (strongly disagree) to 5 (strongly agree) to  ``agree," ``neither", and ``disagree."  


Reflecting the amount of emphasis the participants reported placing on the account as opposed to the message, results regarding what qualities about an account are important to consider when choosing to retweet a message indicated less agreement and certainty. The criterion that seemed most important overall was that the account is ``credible." Our survey did not ask people about how they judge whether or not an account is credible and whether or not that is truly the case. However, this finding suggests that Twitter users do care about whether or not a given account appears trustworthy.


\subsubsection{On the disclaimer ``Retweeting does not mean endorsement''}

In light of the previous findings, of particular interest is the statement encountered in certain Twitter users' profile summary that retweeting does not mean endorsement or agreement. It seems that this statement contradicts the findings presented in this section, so we examined it in greater depth. 

There are relatively few users who include such statement. Using the Twitter Search interface for the terms ``RT, endorsement'' we found 
384 accounts containing both of these terms. An exhaustive categorization of these account showed that 197 (52.7\%) of them belong to people in the Media (reporters, journalists, media producers) including 11 bloggers,  99 (26.5\%) to politicians, 43 (11.5\%) to non-media related companies, 24 (6.4\%) to people who list politics in their interests, and 9 (2.4\%) to academicians. The remaining 0.5\% of accounts were from people indicating that their RT is, in fact, endorsement.

We then tried to collect a comprehensive sample of such profile statements. Using the Twitter API we collected all the user profiles that make a statement along these lines. In fact we found that the terms ``retweet'', ``RT'' co-occur with the terms ``agreement'' or ``endorsement'' in 2,585 profiles. We randomly sampled uniformly  10\% of them and found that media-related accounts comprised 52.65\% and 12.65\% to politicians. 

There is an over-representation of journalists, reporters and media producers who include this disclaimer in their profiles. Other journalists have challenged this practice\footnote{E.g., see Commentaries at the Washington Post http://wapo.st/19OIcfy and NPR's ``Retweets are endorsements'' http://bit.ly/YJBQgQ and http://bit.ly/1v9lBY3} since its inclusion in one's profile is not an effective disclaimer (many people will likely miss it). 

Finally, one has to consider the reason for people placing such a clarification statement in their profile as addressing a concern that one's retweeting practices may be mistaken by others.  
It is reasonable to argue that the mere fact of including this disclaimer is an implicit admission by those using it that others may mistake their intentions,
because for most people retweeting {\em is} endorsement.
Journalists and politicians may worry about the effect of their retweets on their curated public images.


\subsection{Reporters}
\label{reporters}

As we mentioned, several journalists on Twitter include in their Twitter profile description a disclaimer stating that ``retweeting does not equal endorsement." It's almost as if this subgroup of individuals feel a need to distinguish themselves from the attitudes and behavior of other general users on Twitter. To see if participants in the study responded in ways that reflected this distinction, we ran five Welch Two Sample t-tests, as well as the Wilcoxon rank sum tests, to confirm the t-tests. While the t-test's assumption of normality is fairly robust, we wanted to confirm that we would generate similar results even if that assumption is not made.

There was a significant difference in the degree to which reporters (M= 3.4, SD=1.1) versus non-reporters (M= 4.1, SD=0.9) agreed with the statement that {\em ``When retweeting, it is important that the message is something that I agree with,"} t(32.3)=-3, p=0.006. We also found that 57\% of reporters agreed and 17.9\% disagreed with the statement, compared to 75\% of non-reporters agreeing and 3.8\% disagreeing with the statement.


There was also a significant difference in the degree to which reporters (M= 3.3, SD=1.1) versus non-reporters (M=3.9, SD=0.9) agreed with the statement that {\em ``When retweeting, it is important that the message is something that I endorse,"} t(32.7)=-2.9, p=0.007. While 57.1\% of reporters agreed and 21.4\% disagreed with the statement, 69.5\% of non-reporters agreed with the statement and 6.6\% disagreed.


Finally, there was some difference in the degree to which reporters (M= 3.7, SD=1.0) versus non-reporters (M= 4.0, SD=0.9) agreed with the statement that {\em ``When retweeting, it is important that the account is one that I find credible,"} t(35)=-1.88, p=0.07. While 72.4\% of reporters agreed and 13.8\% disagreed with the statement, 76.2\% of non-reporters agreed with the statement and 5.7\% disagreed.


\subsection{Users interested in Politics}


There was a significant difference in the degree to which political users (M= 3.6, SD=1.1) versus non-political users (M= 3.1, SD=1.1) on Twitter find it important to follow accounts that share their opinion, t(116)=2.7, p=0.008. While 76.1\% of political users were motivated and 4.5\% believed that when choosing to follow certain accounts it was important that the accounts share their opinions, 71.6\% of non-political users believed it important and 5.9\% did not. The test for the  difference in the amount by which political users  versus non-political users on Twitter tend to retweet accounts that they don't follow was not significant.



\section{A Meta-analysis of research papers}
\label{meta-analysis}

To focus our research question a bit further, we did a meta-analysis of the research corpus that has been published in the last several years on Twitter. 
Starting with abstract inspection, we collected over 100 relevant papers published since 2008 and 2013 in three major conference venues related to the subject:  AAAI ICWSM, IEEE SocialCom, and WWW. From the initial core we expanded the coverage to references found in the papers from these three venues.
The page limit of this conference does not allow us to expand this section here. We  simply state that our findings indicate support in the survey results, and provide insight on how the emotions associated with hashtags can affect the retweeting rates and visibility of a topic being discussed on Twitter.

\section{Conclusions}
\label{conclusions}

While in the past there have been several claims in the literature about possible reasons for retweeting, no conclusive study of why people retweet appears. This paper aims to close this gap.
Employing a user survey, a study of user profiles, and a meta-analysis of over 100 research publications from three relevant major conferences, we answer the question of what users mean when they retweet.
Our findings show that retweeting indicates, not only interest in a message, but also trust in the message and the originator, and agreement with the message contents. However, the findings are significantly different for journalists,
who are more likely to include a disclaimer on their user profiles that their retweeting does not mean agreement or endorsement.
The inclusion of hashtags strengthens the signal of agreement, especially when the hashtags are related to politics.

We believe that our findings are an important contribution to social media research and to understanding of how some journalists treat Twitter. 
In particular, it can help make sense of the crowd's belief of the validity of a rumor spreading \cite{CJ14-TRAILS} (see {\tt http://bit.ly/twittertrails}).


\section{Acknowledgements}
This research was partially supported by NSF grant CNS-1117693 and by the Wellesley Science Trustees Fund. The authors would like to thank Prof. Jonathan Cheek and Prof. Robin Akert for advise designing the survey.








\end{document}